\documentclass{custom-template}
\usepackage[maxnames=6,minnames=1]{biblatex}
\usepackage[colorlinks,urlcolor=blue,linkcolor=blue,citecolor=blue]{hyperref}
\expandafter\def\expandafter\UrlBreaks\expandafter{\UrlBreaks\do\/\do\*\do\-\do\~\do\'\do\"\do\-}
\usepackage{upmath,color}

\begin{document}

\title{In-Network Collective Operations: \\Game Changer or Challenge for AI Workloads?}

\author{Torsten Hoefler}
\affil{ETH Zürich, Zurich, Switzerland and Microsoft, Redmond, USA}

\author{Mikhail Khalilov}
\affil{ETH Zürich, Zurich, Switzerland}

\author{Josiah Clark}
\affil{AMD, Santa Clara, USA}

\author{Surendra Anubolu}
\affil{Broadcom Inc., San Jose, USA}

\author{Mohan Kalkunte}
\affil{Broadcom Inc., San Jose, USA}

\author{Karen Schramm}
\affil{Broadcom Inc., San Jose, USA}

\author{Eric Spada}
\affil{Broadcom Inc., San Jose, USA}

\author{Duncan Roweth}
\affil{Hewlett Packard Enterprise, Palo Alto, USA}

\author{Keith Underwood}
\affil{Hewlett Packard Enterprise, Palo Alto, USA}

\author{Adrian Caulfield}
\affil{Microsoft, Redmond, USA}

\author{Abdul Kabbani}
\affil{Microsoft, Redmond, USA}

\author{Amirreza Rastegari}
\affil{Microsoft, Redmond, USA}

\markboth{In-Network Collective Operations: Game Changer or Challenge for AI Workloads?}{In-Network Collective Operations: Game Changer or Challenge for AI Workloads?}

\begin{abstract}\looseness-1This paper summarizes the opportunities of in-network collective operations (INC) for accelerated collective operations in AI workloads. We provide sufficient detail to make this important field accessible to non-experts in AI or networking, fostering a connection between these communities. Consider two types of INC: Edge-INC, where the system is implemented at the node level, and Core-INC, where the system is embedded within network switches. We outline the potential performance benefits as well as six key obstacles in the context of both Edge-INC and Core-INC that may hinder their adoption. Finally, we present a set of predictions for the future development and application of INC.
\end{abstract}

\maketitle

\chapteri{M}odern artificial intelligence (AI) and High-Performance Computing (HPC) workloads necessitate substantial computational resources, particularly when dealing with Large Language Models (LLMs). The size and complexity of these models render single accelerators, such as graphical processing units (GPUs), insufficient for efficient training and inference. Consequently, distributed systems have become integral to the training, serving, and inference processes of LLMs.

To address these computational demands, organizations are deploying some of the most powerful supercomputers, equipped with tens of thousands of GPUs. These supercomputers facilitate the extensive parallel processing required for LLM training. The distribution of LLMs primarily uses three fundamental dimensions of parallelism: Data Parallelism (DP), Pipeline Parallelism (PP), and Tensor Parallelism (TP)~\cite{cite-dl-survey}. All those methods use collective operations that may be accelerated with special "In-Network Collective" (INC) hardware support. We proceed to first explain their potential use for AI model training and inference and then discuss their benefits and complexities.

\begin{figure*}[ht]
\centerline{\includegraphics[width=38pc]{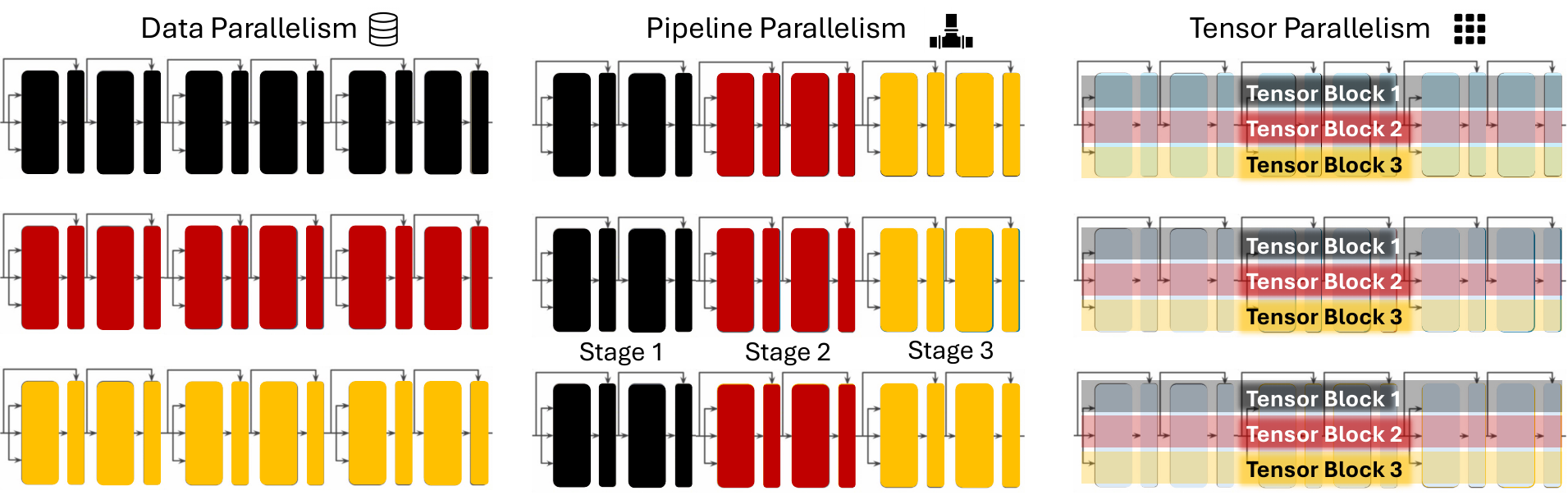}}
\caption{Three dimensions of AI parallelism. Colors mark three different accelerators, potentially 27 total when combined.}\vspace*{-5pt}
\label{fig:ai_parallelism}
\end{figure*}

\begin{figure*}[ht]
\centerline{\includegraphics[width=38pc]{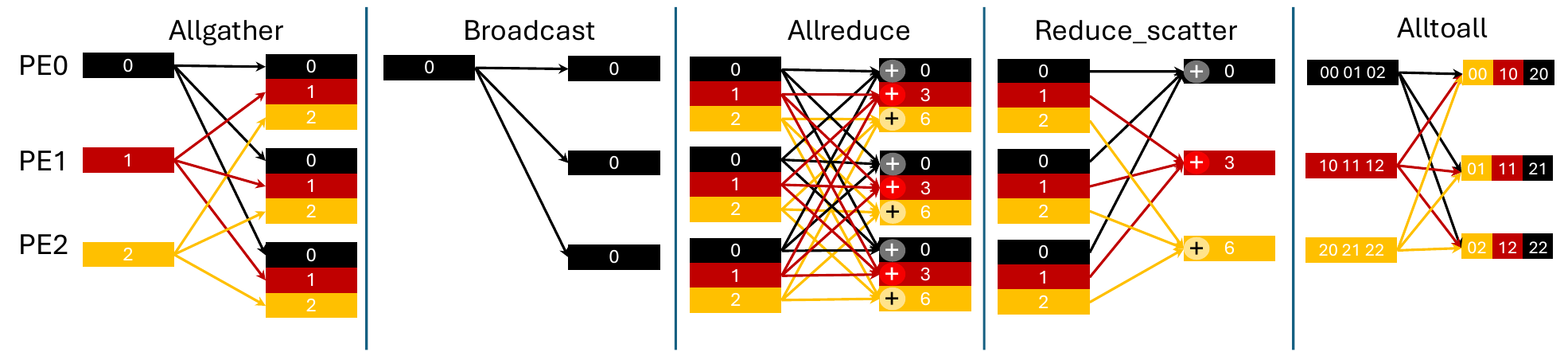}}
\caption{Most frequently used collective operations illustrated with an example involving three processes.}\vspace*{-5pt}
\label{fig:collectives}
\end{figure*}

Data Parallelism (DP) is used during training by replicating model copies to accelerate training by distributing mini-batches of data across multiple accelerators. Each accelerator processes different subsets of the data simultaneously, thereby enhancing training efficiency. Fully Sharded Data Parallelism (FSDP)~\cite{zero, fsdp} optimizes memory usage of DP by partitioning the model parameters, optimizer state, and gradients across different accelerators, thereby preventing an increase in memory requirements.  

Pipeline Parallelism (PP) is used in inference and training where it distributes the model weights across multiple accelerators, effectively partitioning the model to fit within the local memory of each accelerator. This approach executes different segments of the model pipeline on distinct accelerators, thereby optimizing memory usage and computational efficiency. Yet, usually, the processing time for each batch is increased when going through the pipeline. 

Tensor Parallelism (TP) (or Operator Parallelism) is used in inference and training where it accelerates the processing of individual pipeline stages by distributing the tensors (data arrays) across multiple accelerators. This method divides the tensors into smaller chunks that can be processed in parallel, thus reducing the overall computation time required for each stage of the model.

Figure~\ref{fig:ai_parallelism} illustrates how the three forms of parallelism can be used to distribute training of a transformer model (left to right, three replicas) with three decoder blocks, each consisting of the typical Multi-Head Attention, Normalization, and Feed-Forward layers. Data parallelism shows how the whole model is replicated and pipeline and tensor parallelism show the sharding of the parameters layer-wise, and tensor-block-wise, respectively. Each color mark the distribution to three different accelerators in each example. 

While other forms of parallelism, such as Sequence Parallelism (SP), exist, they exhibit similarities to the aforementioned dimensions and are omitted for simplicity. When distributing models, it is crucial to consider the data distribution and resulting communication required for forward and backward passes of the model. Efficient distribution strategies ensure that data is partitioned and synchronized across all accelerators, minimizing communication overhead and maximizing computational throughput.

\section{The Role of Collective Operations and CCLs in AI}

Collective Communication Operations are used for inter-processor communication during AI training and inference.  These nonblocking operations are defined in the Message Passing Interface (MPI) Standard~\cite{nbc}. Subsets of these operations specific to different GPU architectures are found in processor element (PE) specific Collective Communication Libaries (CCL)~\footnote{E.g., NVIDIA Collective Communications Library (NCCL)}.

MPI-like collective operations form the backbone of most distributed deep learning frameworks. Figure~\ref{fig:collectives} shows the data distribution for the most important collective operations on three accelerators (processing elements, PEs). In Allgather, each processing element (PE) starts with a scalar and \textit{broadcasts} it to each other PE. Broadcast collective is a subset of Allgather. Allreduce starts with a vector at each PE and ends with the sum of the vectors at each PE. Reduce\_scatter performs the same summation but shards the result such that each PE ends up with a single scalar. Alltoall transposes a distributed matrix.

DP and TP use Allreduce for tasks like gradient averaging and distributed tensor processing (e.g., matrix multiplication). FSDP uses Allgather and Reduce\_scatter operations to distribute model parameters, optimizer states, and gradients across multiple accelerators. CCLs support the MPI concept of communicators that defines subsets of communicating groups of processes to perform collective operations. This feature is crucial for AI workloads and INC implementations, because collective operations typically use subsets of the processes in a job, grouped into communicators, as defined by the parallelism schemes outlined above. 

The Mixture of Experts model, as exemplified by DeepSeek-V3~\cite{dsv3}, relies on Alltoall(v) operations to distribute data among experts. While other distribution schemes might necessitate Broadcast operations, such rooted collectives are relatively rare. They are often implemented as Allgather operations, effectively functioning as an all-Broadcast. Additionally, Scan operations could be beneficial for workloads that utilize small neural networks and operate under stringent latency constraints. However, there are no widely known use cases yet for MPI’s Scan or Exscan in this context.

\section{Opportunities for In-Network Computation to Accelerate AI}

\textbf{Edge-INC} offloads some of the CCL collective operations from the compute accelerator to the local network (“edge”) interfaces. This approach reduces memory contention and overall latency at the endpoint by enabling data to be forwarded directly in a streaming manner~\cite{spin}, without the need to be stored in the accelerator's memory. Consequently, this method enhances efficiency and minimizes the performance bottlenecks typically associated with memory access.

\noindent
\textbf{Core-INC} offloads some collective operations into the network (“core”) switches themselves, as exemplified by NVIDIA's SHARP~\cite{sharp}. Switches actively participate in the operation, performing computations such as summing data. This approach leverages the computational capabilities embedded within the network infrastructure, thereby optimizing data processing and reducing latency or bandwidth consumption within the network core. By utilizing Core-INC, the network can achieve significant performance improvements through distributed computation at the network core.

\subsection{Collective Acceleration using Edge-INC}

Edge-INC exclusively involves the host network interface (NI). Approaches such as Portals 4~\cite{p4off} and sPIN~\cite{spin} enable this method and have a minimal implementation footprint, exemplified by techniques such as triggered operations~\cite{p4.3}. Additionally, they can implement advanced communication protocols, such as multicast-based constant-time Broadcast~\cite{bcast}, to further enhance data transfer efficiency and reduce latency. Using specific NI offloads, these methods minimize the computational burden on the host and optimize overall system performance.

Asynchronous progression and full collective operation offload from the accelerator to the NI are enabled, allowing for the complete overlap of computation on the accelerator with communication in the network. Edge-INC eliminates host memory access latency and bandwidth demands by directly forwarding messages from the NI. It simultaneously reduces control overhead on the host while saving significant latency for large-scale systems.

\begin{figure}
\centerline{\includegraphics[width=18.5pc]{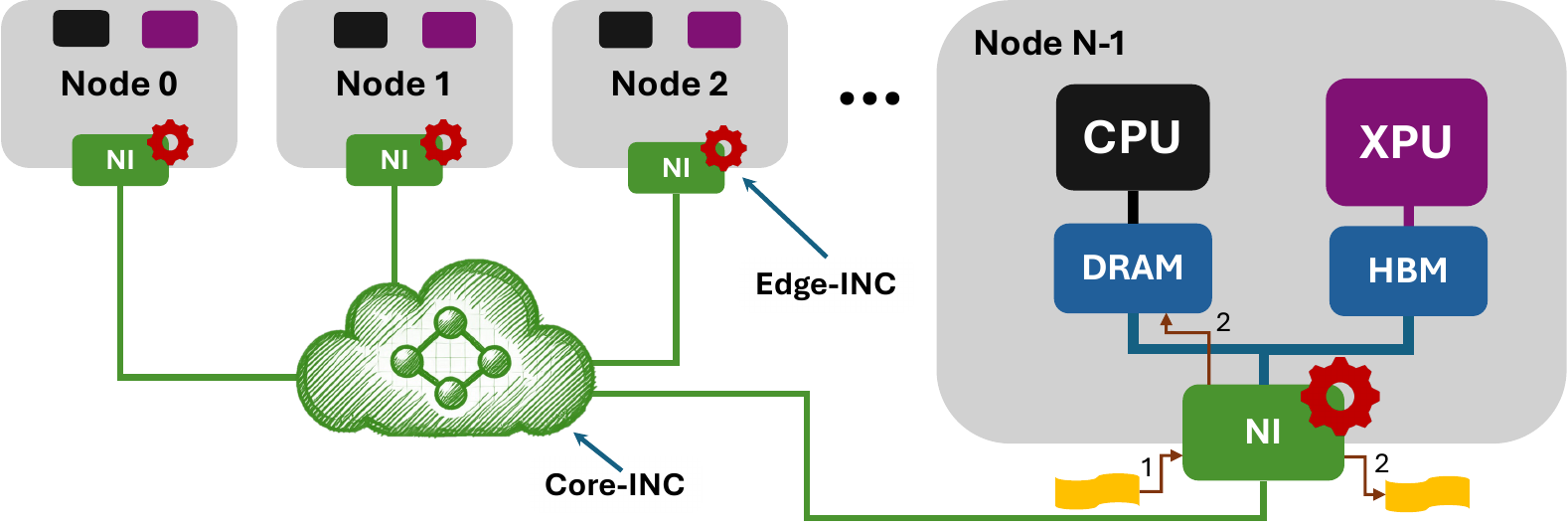}}
\caption{A hybrid system with Edge-INC and Core-INC.}\vspace*{-5pt}
\label{fig:edge_vs_core}
\end{figure}

Figure~\ref{fig:edge_vs_core} illustrates an Edge-INC scenario with N nodes, each equipped with an accelerated NI, depicted by the red cog. It also annotates where Core-INC would live. In the context of a pipelined linear Broadcast or the second phase of a ring reduction, node i receives data from node $i-1$ (modulo N), deposits it into the main memory, and subsequently sends it to node $i+1$ (modulo N). The figure demonstrates a yellow packet arriving at step 1, being deposited into the main memory and simultaneously forwarded at step 2 by the NIC.

\begin{figure*}
\centerline{\includegraphics[width=38pc]{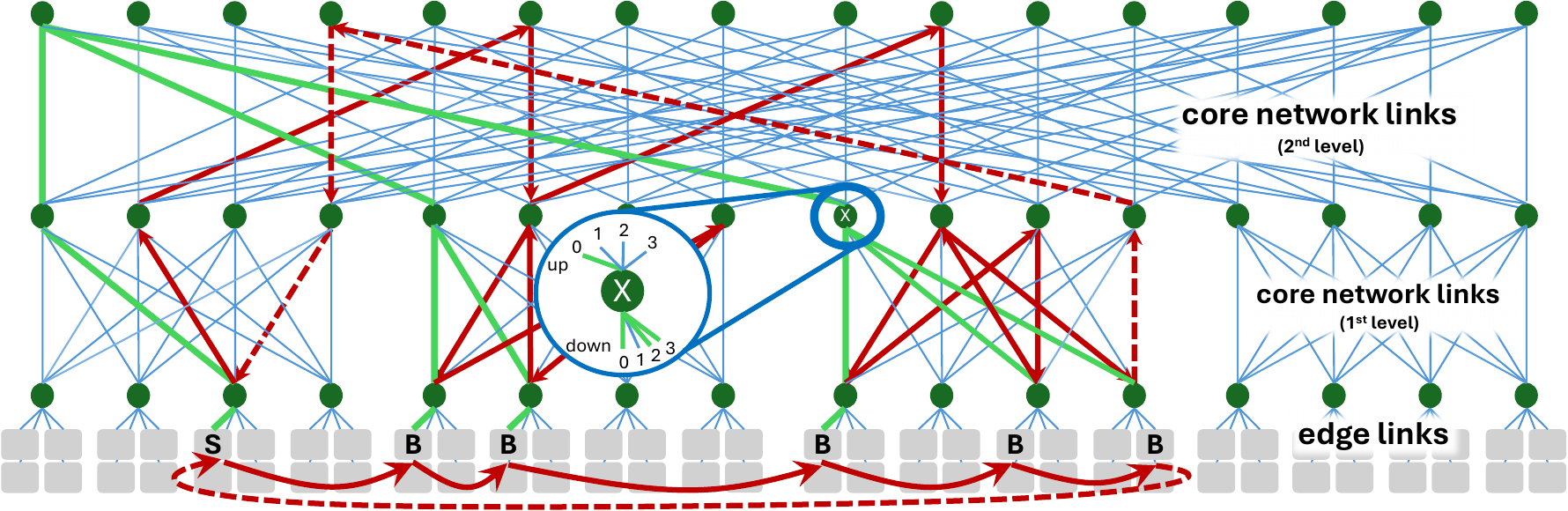}}
\caption{The INC-based Allreduce algorithm requires less total data movement in comparison to the ring algorithm.}\vspace*{-5pt}
\label{fig:inc_network}
\end{figure*}

Without Edge-INC, the NI would first write this data into DRAM and then read it again to transmit it, resulting in an additional memory transaction. This process essentially doubles the load on the host bus and the CPU or accelerator memory (illustrated in blue). For high-bandwidth NICs, this increased load can lead to significant computational slowdowns.

\subsection{Collective Acceleration using Core-INC}

While Edge-INC reduces node-level contention and latencies, Core-INC  reduces communication costs at the network level. However, not all collective operations can be equally accelerated by Core-INC. \textbf{Core-INC is most effective when data is reduced} during the operation, as in (All)Reduce, \textbf{or replicated} during the operation, as in Broadcast and Allgather.

\textbf{Broadcast} operations benefit from various bandwidth- and latency-optimal algorithms tailored for endpoint-based implementations. For instance, pipelined ring algorithms are optimal for handling very large (infinitely large) messages, while Fibonacci trees are most efficient for the smallest messages, effectively covering the range in between~\cite{colls}. The trade-off between bandwidth and latency is particularly noteworthy; Core-INC can leverage switch-reliable multicast to significantly reduce latency during Broadcast operations. 

\textbf{Allreduce} operations benefit most from the network switches performing the operations. Core-INC also cuts the needed network bandwidth in half~\cite{sc24}. Endpoint-based algorithms must send and receive each segment twice, once in the reduction and once in the broadcast phase, but Core-INC only requires sending it once and receiving it once because it is reduced by the switches.

Figure~\ref{fig:inc_network} shows the benefits of a Core-INC based Allreduce compared to a endpoint-bandwidth-optimal ring algorithm. The figure shows a full fat tree and a subset of nodes marked “B” that participate in the Allreduce. Core-INC communicates along a single tree (shown in green) where all nodes send to the root at the top-left switch in the network and then the root reliably broadcasts the result back to all involved nodes. Here, each node sends each segment once and then receives the final result. During the reduction phase, the highlighted switch “X” collects the input from its three incoming children, reduces (e.g., sums) it, and sends it towards the root switch. During the broadcast phase, switch X replicates the data from the root switch to all its green children. The red schedule shows an endpoint-based ring algorithm and possible routes through the network. This algorithm, albeit occupying the same number of links, has to perform two rounds.

\textbf{Allgather} is more complex as the data is not reduced in the network. Yet, it can save network bandwidth because Allgather is an all-Broadcast where each node broadcasts to all other nodes. Node S in the figure above would send its data to the root switch, which would replicate it twice to its children, who would then replicate it again to their children who in turn will deliver it to the destination nodes. Excluding edge links, the data traverses only 9 links. The red schedule excluding the dashed lines shows a standard pipelined Broadcast occupying a total of 13 links. Larger jobs and oversubscribed networks would show even more pronounced benefits. Thus, Core-INC can reduce the network bandwidth utilization significantly~\cite{sc24}. 

\textbf{Reduce\_scatter} can be implemented as multiple reductions that are identical to broadcast trees. Here, the same idea as Allgather applies and bandwidth can be saved in the core network. Furthermore, concurrent Reduce\_scatter can be combined with Allgather in Core-INC to gain bandwidth savings up to $2\times$ due to their different bottlenecks~\cite{sc24}. 

\textbf{Alltoall} cannot be optimized easily with Core-INC as there is no reduction in data at all. Alltoall simply transposes a large array. Here, Edge- or Core-INC could be used to synchronize nodes to orchestrate congestion-free schedules for sending the data. This is generally hard given that the network may not be exclusively used by a single tenant. 

Core-INC and Edge-INC both have complex relationships with the system itself. From a reliability perspective, Core-INC utilizes fewer links, but becomes more dependent on the links it uses and carries state in the switches that is not resilient to switch failure. Edge-INC can use traditional techniques to route around failures, but incurs delays during failures and is more dependent on a high fraction of the bandwidth being available. Job fragmentation can cause point-to-point communications in Edge-INC to take substantially longer, whereas Core-INC is more immune to fragmentation - until the fragmentation reaches a level where the network is no longer able to achieve a data reduction. Both Edge-INC and Core-INC can be complementary and multiply their benefits. For example, a local NI can take charge of coordinating the Core-INC such that the accelerator is completely freed from communication overheads and full overlap of communication and computation can be achieved. 

\section{Problems for In-Network Compute Acceleration of AI}

The advantages of In-Network Computation (INC) are both evident and substantial, with potential traffic reductions at both edge links and in the core network of up to $2\times$ for operations such as Allreduce, Reduce\_scatter, Broadcast, and Allgather. Additionally, INC can significantly reduce host memory load and provide opportunities for overlapping computation and communication during collective operations~\cite{nbc}. However, the intricacies of INC can be complex and challenging to navigate. In the following sections, we present several obstacles that architects and engineers developing INC systems must consider.

\subsection{Low Precision Data Types}
One of the most effective optimizations in deep learning and AI is the utilization of low-precision data types. Reducing the number of bits used to represent numerical values not only linearly decreases data volume and movement but can also lead to a quadratic increase in computational performance. Specifically, reducing values from $16$-bit to $8$-bit precision results in a $2\times$ savings in memory and memory bandwidth, as well as a potential $4\times$ speedup in computations. The computational speedup is because the multiplication of n-bit integers generally requires $O(n^2)$ logic or time, making low-precision representations highly advantageous for efficient AI computations.

Low-precision types can only express a narrower range of numbers. An $8$-bit integer can represent $256$ distinct values, while a $4$-bit integer is limited to just $16$ different values. The selection of which numbers these types represent can be determined either algorithmically, such as setting a uniform range like $-128$ to $127$ for signed integers, or through a predefined code-book. Moreover, the range can be dynamically adjusted by scaling factors that are applied block- or tensor-wise to better fit the data's dynamic range. However, a significant challenge with low-bit representations, particularly in processing long sequences, is temporary overflow, underflow, or accumulating rounding errors. This occurs when intermediate computations exceed the representational capacity of the low-bit format, even if the final result could theoretically be accurately represented within that format.

To illustrate the challenges with low-precision arithmetic, consider performing a series of operations using a signed int4 type (range $-8$ to $7$): $7-5+5+5-3-7$. If we compute from left to right, we get intermediate results of $7$, $2$, $7$, then an overflow to $-3$, $-6$, and an underflow to $2$. Although the final result might be correct, these intermediate results are inaccurate, leading to errors when used in further calculations like dot-products or matrix multiplications, especially when combined with multiplication operations. Another example is the multiplication task $2*2*3/2$, which yields intermediate results of $2$, $4$, then an overflow to $-4$, and finally the incorrect $-2$, instead of the expected $6$. These precision issues also affect sequences of mixed addition and multiplications. Since floating-point operations fundamentally involve both multiplication and addition in operations (e.g., floating point multiplication multiplies the mantissas and adds the exponents), the same problems can occur. To prevent these errors, AI accelerators employ higher-precision internal accumulation registers.

Core-INC requires sending intermediate results to upstream switches by design, which introduces challenges with precision and bandwidth. Since the maximum bandwidth savings from Core-INC is a factor of two, transmitting numbers with higher precision to maintain accuracy would essentially negate this advantage. Compounding the issue, most internal higher-precision registers are significantly larger than just twice the size of the input data types. This situation is commonly referred to as the "problem of communicating the accumulator" in Core-INC systems, where accumulators need to be relayed between switches. Conversely, in Edge-INC scenarios, where large vectors of numbers are reduced, one can mitigate this by designating a specific host for each range, thereby allowing for the local maintenance of a high-precision accumulator at each Network Interface (NI), keeping the benefits of precision without the overhead of excessive data transmission.

One potential, though complex, workaround involves sending only low-precision values from the edge hosts to the first reduction switch, then using a high-precision accumulator for communication between core switches, finally casting down to the lower target precision at the root switch. This approach requires careful consideration because the accumulation switches are not always the first in the chain. For instance, in the Core-INC example figure above, the source S does not directly connect to an accumulation switch; instead, the accumulator size should increase at the second hop, which coincidentally is also the root. In other parts of the green network tree in our example, the precision would need to be increased at the second switch encountered. If more than two group members would be connected to an edge switch, then the result would need to be upcast there. Even if this strategy is implemented correctly, it still necessitates transmitting at least double the data volume upwards through intermediate links, thus diminishing the potential for bandwidth savings within the core network. Furthermore, such savings would depend strongly on the tree topology.

\subsection{Vector Data Types}

Some deep learning accelerators and workloads leverage vector data types, which, like the quantized numbers previously discussed, incorporate scaling factors. However, these scaling factors are not uniformly applied across entire tensors or large blocks but are instead tailored to smaller blocks that constitute the basic units of computation. An example of this is a set of 16 integer values scaled by an exponential floating-point value, a method referred to as block floating point. This technique has been adopted to formulate various blocked data types, such as MxFP~\cite{mxfp}, enhancing efficiency in deep learning computations.

If block floating-point and similar vector data types are extensively utilized, INC systems might need to inherently support these formats. Although one could convert these block formats into types that INC currently supports, such conversions introduce additional overhead. This overhead could diminish the performance benefits of INC, particularly since operations on these vector types can be executed very efficiently on native architectures using traditional collective algorithms like ring. Consequently, implementing complex type handling within the INC switch or Network Interface (NI) might be necessary, which could increase both cost and design complexity.

Another broad challenge in this domain is the rapid evolution of data types used in deep learning. Over the past decade, we've witnessed a "Cambrian explosion" of various data types, with new formats like BF16, E4M3, and E5M2 gaining quick and widespread acceptance due to their availability on modern accelerators. This swift pace of change in data types poses a significant challenge for the networking field, which often operates within a slower silicon design cycle, making it difficult to keep pace with the latest advancements in deep learning.

\subsection{Sparse Vector Reductions}

Sparse computations offer significant advantages for deep learning workloads by reducing computational overhead and memory usage~\cite{sp-survey}. This advantage extends into network computations but introduces the problem of fill-in. Similar to the issue of communicating accumulators, fill-in can lead to even more substantial slowdowns. The core problem arises when multiplying two sparse vectors in a large index space; the result often becomes much larger because it must accommodate as many elements as the union of the indices from both vectors. If non-zeros are randomly distributed, this leads to a rapid increase in vector size as computations progress through the reduction tree. Consequently, vectors might become so dense that even switching to a dense representation becomes more efficient at some height along the tree~\cite{cedric, daniele}.

Supporting sparse reductions in Core-INC is inherently complex and prone to errors, which can negate much of the efficiency gains provided. To circumvent this issue, one potential strategy involves sharding the index space across endpoints, thereby directing all values associated with a particular index space to their respective endpoint. It's mathematically provable, given certain distributions of non-zero elements, that such an approach can optimize communication volume, potentially reducing the data traffic in the core network more effectively than with Core-INC methods. These advanced, endpoint-based schemes~\cite{shigang} could be effectively realized through Edge-INC, allowing for all the associated benefits to be fully exploited.

\subsection{Bitwise Reduction Result Reproducibility}

For certain workloads and use-cases, such as debugging, bitwise reproducibility is critical. This is straightforward with integer data types, but floating-point sum is not associative. Therefore, floating-point calculations can only achieve bitwise reproducibility if the exact order of operations is maintained or if specific reproducibility schemes are employed~\cite{repro-blas, andrea}. However, these schemes typically introduce up to twice the data and computational overhead, which can negate the $2\times$ efficiency gain offered by Core-INC, akin to the problem encountered with accumulator communication. While there might be room for innovation in developing schemes for INC, ensuring a consistent order of operations is often the simplest solution.

Users care about  intra-job and inter-job reproducibility. Intra-job reproducibility ensures that reductions within a single job are bitwise identical, while inter-job reproducibility extends this guarantee across different jobs, specifically to those involving the same number of processes. Achieving inter-job reproducibility necessitates ensuring an identical tree structure for data reduction across jobs, which presents significant challenges because the distribution of processes across nodes can vary greatly between different job executions. Ensuring consistent tree ordering is both theoretically and practically complex and is an open research problem. It might not be feasible for certain configurations of process-to-node mappings.

\subsection{Endpoint Interfaces and Coordination}

When implementing INC, it's essential for the switch to incorporate a basic networking stack to handle reliability, flow control, and congestion management. This is due to the shared nature of the physical link, where both INC and regular end-to-end traffic compete for bandwidth, necessitating compatible congestion control mechanisms. Additionally, NIs at the endpoints must manage INC states and contexts in hardware, leading to increased complexity in logic (e.g., when considering link aggregation) and additional memory overhead.

Constructing Core-INC groups and their corresponding reduction trees presents significant system-level challenges and hindered early adoption. The hardware resources available at each switch constrain the number of trees it can support. Given that switches are interconnected in complex topologies, and jobs launch on dynamic subsets of nodes, creating and dismantling these trees must be managed efficiently at job (or even communicator) initiation and termination. Multi-tenancy can exacerbate this by increasing the number of trees within the network, potentially necessitating strict isolation measures. However, many of these complexities can be sidestepped with Edge-INC systems, which can operate effectively using simpler switches designed purely for data movement, thereby reducing the overhead associated with managing intricate network structures.

\subsection{Encryption and Authentication}

Encryption and authentication crucial in confidential computing systems. Unfortunately, Core-INC faces significant challenges in supporting end-to-end encryption due to its data manipulation operations. While homomorphic encryption presents a potential solution, its application is currently limited and most effective for integer data types only. Chrapek et al. develop several initial ideas for Homomorphic Core-INC systems~\cite{marcin}. Yet, for a system to support all operations and data types, it would necessitate extending trust to the switches, thereby substantially expanding the trust domain. Employing separate keys for INC communication can help limit the trust domain's scope. However, any data transmitted through INC remains susceptible to compromise.

Implementing encryption and authentication in a Core-INC system remains challenginga and a topic for research, even when utilizing separate keys and security domains. This complexity arises because switches must partake in key rotation and re-keying, potentially incorporating key derivation and other advanced security mechanisms. The necessary overhead in memory for managing these keys and the logic required for system administration is not only intricate and costly but also introduces additional security vulnerabilities. Edge-INC systems may simplify the security handling as only the local NI would need to be part of the trust domain. 

\section{Performance for INC in Deep Learning}

The primary objective of INC is to expedite computations within the network and decrease communication volumes. While reducing communication volumes is advantageous because it frees network resources to handle other types of traffic, accelerating computations presents a more intricate challenge. In the context of accelerating real-world applications, a variant of Amdahl's Law becomes a significant constraint. Consequently, claims of a tenfold improvement in application performance should prompt scrutiny from astute performance experts.

\begin{figure}
\centerline{\includegraphics[width=18.5pc]{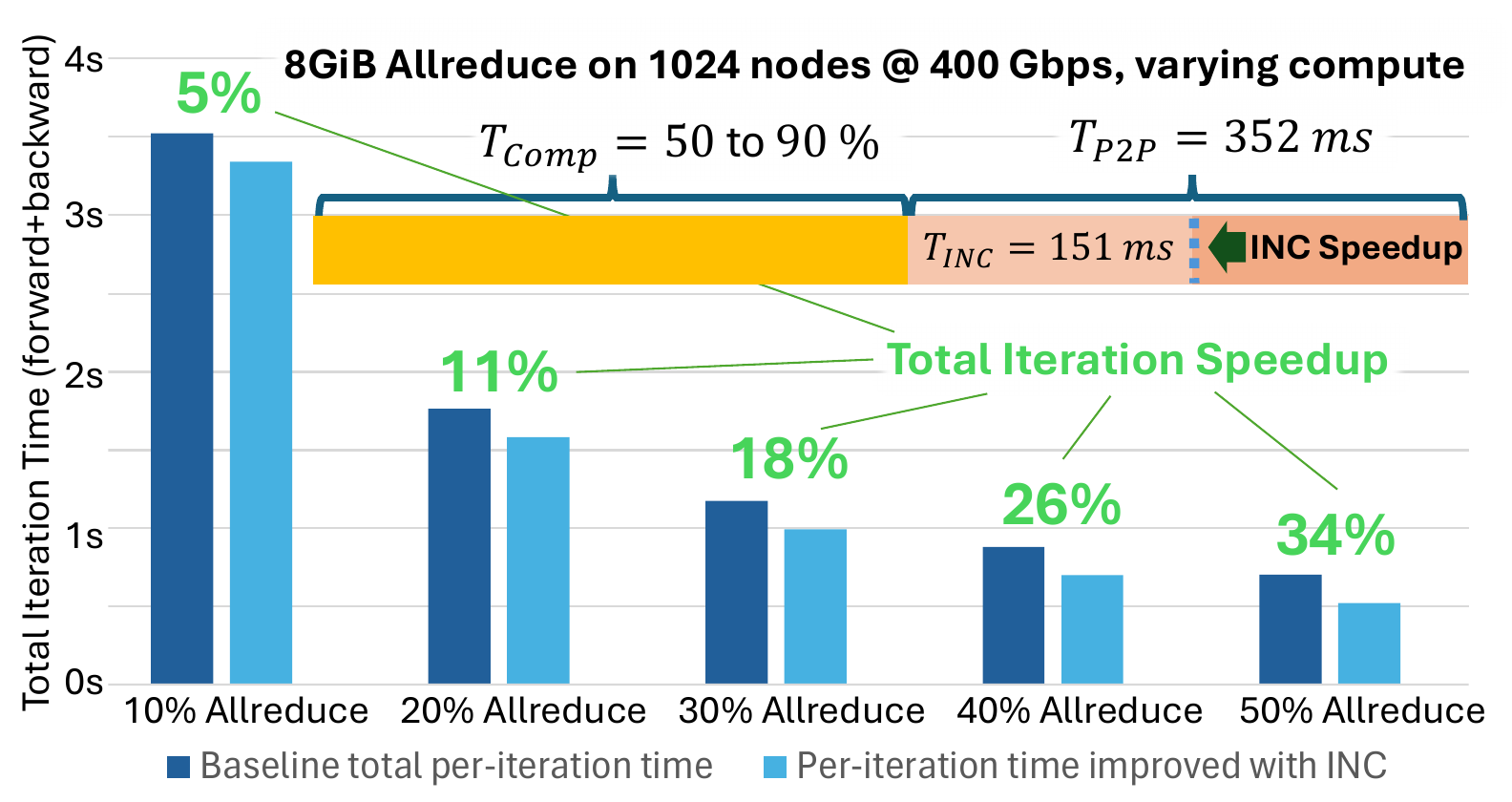}}
\caption{Performance model for INC-enabled DP training.}\vspace*{-5pt}
\label{fig:dp_gains_analysis}
\end{figure}

In Figure~\ref{fig:dp_gains_analysis} we consider the following simple example of DP parallel training where we model a fixed 8GiB Allreduce with varying iteration times to adjust the communication overhead (portion of iteration time (\%) spent in Allreduce) between $10$-$50$\% on the x-axis. Without INC, the ring Allreduce takes 352ms and INC can reduce the time to 151ms, a nearly 60\% speedup. Yet, due to Amdahl’s law, the maximum speedup achieved is $34$\%. In a realistic case where non-overlapped communication overheads are $20$\%, INC would be limited to an $11$\% overall speedup.  We note that even a $5$\% reduction in runtime due to INC can be substantial given that the network typically costs less than 20\% of the system.

\section{Summary and Predictions}

We highlight numerous potential benefits that In-Network Computing (INC) can offer to collective operations driving AI inference and training. However, we also identify several major challenges specific to AI workloads that any INC system must address. We predict that these complexities and the potentially limited advantages may result in slow adoption. Consequently, we foresee that the adoption of INC will remain gradual for the next years.

A promising area for the adoption of In-Network Computing (INC) is within a local network context, where communication costs are particularly high. In such environments, single-switch solutions can provide significant benefits. These solutions would be simpler to engineer compared to more complex multi-switch deployments, as they can essentially function as extensions to the existing network nodes. By leveraging single-switch INC solutions, it becomes feasible to enhance performance and reduce communication overhead without extensive reengineering of the network infrastructure. Given the current advancements in networking technology, we anticipate that these single-switch INC solutions will be adopted in more use-cases soon and will prove to be highly effective in addressing the communication challenges faced by local networks. This approach not only streamlines implementation but also maximizes the performance gains in a cost-effective manner.

Standardization provides a common ground for various products and approaches to make them accessible to wide community and enable fair market competition. Thus, it is crucial for the success and widespread adoption of INC technologies. One promising initiative towards this goal is Ultra Ethernet, spearheaded by a consortium working group that includes many of the co-authors of this paper. The current plans focus on Core-INC with Edge-INC being on the horizon. The development of any INC specification must be lean and straightforward to justify the necessary investments. Ensuring simplicity and clarity will be key to its effectiveness and practical implementation. Although it remains to be seen whether the standard will meet the high expectations set for it, there is a strong prediction that the upcoming specification will significantly benefit the adoption of INC. Nevertheless, it is important to acknowledge that INC will continue to face challenges, particularly in terms of technical complexity and integration into existing systems. The ongoing efforts in standardization and collaboration within the consortium are vital steps towards overcoming these challenges and realizing the potential benefits of INC.

\section{ACKNOWLEDGMENTS}
We thank the whole INC working group in UEC and all others who have contributed to the development of the thinking behind this article!

\def\refname{REFERENCES}

\printbibliography

@article{cite-dl-survey,
  author={Tal Ben-Nun and Torsten Hoefler},
  title={{Demystifying Parallel and Distributed Deep Learning: An In-Depth Concurrency Analysis}},
  journal={ACM Comput. Surv.},
  year={2019},
  publisher={ACM},
}

@inproceedings{zero,
  title={Zero: Memory optimizations toward training trillion parameter models},
  author={Rajbhandari, Samyam and Rasley, Jeff and Ruwase, Olatunji and He, Yuxiong},
  booktitle={SC'20},
  Publisher={IEEE}
}

@article{fsdp,
  title={Pytorch fsdp: experiences on scaling fully sharded data parallel},
  author={Zhao, Yanli and Gu, Andrew and Varma, Rohan and Luo, Liang and Huang, Chien-Chin and Xu, Min and Wright, Less and Shojanazeri, Hamid and Ott, Myle and Shleifer, Sam and others},
  journal={arXiv preprint arXiv:2304.11277},
}

@inproceedings{nbc,
  author={Torsten Hoefler and Andrew Lumsdaine and Wolfgang Rehm},
  title={{Implementation and Performance Analysis of Non-Blocking Collective Operations for MPI}},
  booktitle={SC'07},
  publisher={IEEE, ACM},
}

@article{dsv3,
  title={Deepseek-v3 technical report},
  author={Liu, Aixin and Feng, Bei and Xue, Bing and Wang, Bingxuan and Wu, Bochao and Lu, Chengda and Zhao, Chenggang and Deng, Chengqi and Zhang, Chenyu and Ruan, Chong and others},
  journal={arXiv preprint arXiv:2412.19437},
}

@inproceedings{spin,
  author={Torsten Hoefler and Salvatore Di Girolamo and Konstantin Taranov and R. E. Grant and Ron Brightwell},
  title={{sPIN: High-performance streaming Processing in the Network}},
  booktitle={SC'17},
}

@inproceedings{sharp,
  title={{Scalable hierarchical aggregation protocol (SHArP): A hardware architecture for efficient data reduction}},
  author={Graham, Richard L and Bureddy, Devendar and Lui, Pak and Rosenstock, Hal and Shainer, Gilad and Bloch, Gil and Goldenerg, Dror and Dubman, Mike and Kotchubievsky, Sasha and Koushnir, Vladimir and others},
  booktitle={COMHPC 2016},
  organization={IEEE}
}

@article{p4off,
  author={Salvatore Di Girolamo and P. Jolivet and K. D. Underwood and Torsten Hoefler},
  title={{Exploiting Offload Enabled Network Interfaces}},
  journal={IEEE MICRO},
  year={2016},
  volume={36},
  number={4},
  publisher={IEEE},
  source={http://www.unixer.de/~htor/publications/},
}

@techreport{p4.3,
  title={{The Portals 4.3 Network Programming Interface}},
  author={Brightwell, Ronald and Schonbein, William Whit and Pedretti, Kevin and Hemmert, Karl Scott and Maccabe, Arthur B and Grant, Ryan E and Barrett, Brian W and Underwood, Keith and Riesen, Rolf and Hoefler, Torsten and others},
  institution={Sandia National Lab.(SNL-NM), Albuquerque, NM (United States)}
}

@inproceedings{bcast,
  author={Torsten Hoefler and Christian Siebert and Wolfgang Rehm},
  title={{A practically constant-time MPI Broadcast Algorithm for large-scale InfiniBand Clusters with Multicast}},
  booktitle={CAC'07 Workshop},
  publisher={IEEE},
}

@article{colls,
  author={Torsten Hoefler and D. Moor},
  title={{Energy, Memory, and Runtime Tradeoffs for Implementing Collective Communication Operations}},
  journal={Journal of Supercomputing Frontiers and Innovations},
  year={2014},
  volume={1},
  number={2},
  publisher={SuperFri Open Journal},
}

@inproceedings{sc24,
  author={Mikhail Khalilov and Salvatore Di Girolamo and Marcin Chrapek and Rami Nudelman and Gil Bloch and Torsten Hoefler},
  title={{Network-Offloaded Bandwidth-Optimal Broadcast and Allgather for Distributed AI}},
  booktitle={SC'24},
  publisher={IEEE Press},
}

@article{mxfp,
  title={Microscaling data formats for deep learning},
  author={Rouhani, Bita Darvish and Zhao, Ritchie and More, Ankit and Hall, Mathew and Khodamoradi, Alireza and Deng, Summer and Choudhary, Dhruv and Cornea, Marius and Dellinger, Eric and Denolf, Kristof and others},
  journal={arXiv preprint arXiv:2310.10537},
}

@article{sp-survey,
  author={Torsten Hoefler and Dan Alistarh and Tan Ben-Nun and Nikoli Dryden and Alexandra Peste},
  title={{Sparsity in Deep Learning: Pruning and growth for efficient inference and training in neural networks}},
  journal={Journal of Machine Learning Research},
  year={2021},
  volume={22},
  number={241},
}

@inproceedings{cedric,
  author={Cedric Renggli and Dan Alistarh and Mehdi Aghagolzadeh and Torsten Hoefler},
  title={{SparCML: High-Performance Sparse Communication for Machine Learning}},
  booktitle={SC'19},
  publisher={ACM},
}

@inproceedings{daniele,
  author={Daniele De Sensi and Salvatore Di Girolamo and Saleh Ashkboos and Shigang Li and Torsten Hoefler},
  title={{Flare: Flexible In-Network Allreduce}},
  booktitle={SC'21},
  publisher={ACM},
}

@inproceedings{shigang,
  author={Shigang Li and Torsten Hoefler},
  title={{Near-Optimal Sparse Allreduce for Distributed Deep Learning}},
  booktitle={PPoPP 2022},
}

@inproceedings{andrea,
  author={Andrea Arteaga and Oliver Fuhrer and Torsten Hoefler},
  title={{Designing Bit-Reproducible Portable High-Performance Applications}},
  booktitle={IPDPS 2014},
  publisher={IEEE},
}

@techreport{repro-blas,
    Author= {Demmel, James and Ahrens, Willow and Nguyen, Hong Diep},
    Title= {Efficient Reproducible Floating Point Summation and BLAS},
    Number= {UCB/EECS-2016-121},
}

@inproceedings{marcin,
  author={Marcin Chrapek and Mikhail Khalilov and Torsten Hoefler},
  title={{HEAR: Homomorphically Encrypted Allreduce}},
  booktitle={SC'23},
  publisher={ACM},
}

\begin{IEEEbiography}{Torsten Hoefler} {\,} is a Professor of Computer Science at ETH Zurich and a Scientific Advisor to Microsoft. He is co-chair of the Ultra Ethernet Transport WG, advancing Ethernet for AI \& HPC. Hoefler received his PhD at Indiana University and he is a fellow of IEEE, ACM, and ELLIS. Contact him at htor@ethz.ch.
\end{IEEEbiography}

\begin{IEEEbiography}{Mikhail Khalilov} {\,} is a doctoral student at ETH Zürich. His research focuses on HPC \& datacenter interconnects. Khalilov received his MSc degree in Applied Math and Informatics from HSE University, Moscow. Contact him at mkhalilov@ethz.ch.
\end{IEEEbiography}

\begin{IEEEbiography}{Josiah Clark} {\,} is an ML Engineer at AMD who developed AMD’s initial Core-INC theory and architecture.  His email is josiah.clark@amd.com.
\end{IEEEbiography}

\begin{IEEEbiography}{Surendra Anubolu} {\,} is a Sr Technical Director and a Distinguished Engineer at Broadcom. His email is surendra.anubolu@broadcom.com.
\end{IEEEbiography}

\begin{IEEEbiography}{Mohan Kalkunte} {\,} is the Vice President of Architecture and Technology responsible for the architecture development of switches for Enterprise, Data Center, and Service Provider markets at Broadcom.  He is an IEEE fellow and has over 150 patents. His email is mohan.kalkunte@broadcom.com.
\end{IEEEbiography}

\begin{IEEEbiography}{Karen Schramm} {\,} is a Vice President of Architecture at Broadcom, responsible for Ethernet Network Adapters. She is co-chair of the Ultra Ethernet Transport WG, advancing Ethernet for AI \& HPC. Her email is karen.schramm@broadcom.com.
\end{IEEEbiography}

\begin{IEEEbiography}{Eric Spada} {\,} is a Distinguished Engineer at Broadcom, working on Ethernet NIC architecture.  He is co-chair of the Ultra Ethernet Transport SW working group and editor of UET Transport Security specification. His email is eric.spada@broadcom.com.
\end{IEEEbiography}

\begin{IEEEbiography}{Duncan Roweth} {\,} is an HPE Fellow and the Chief Architect for Slingshot. While at Cray, Duncan was an instigator of the Slingshot program seeing it from early research to deployment on all three of the US exascale systems.
\end{IEEEbiography}

\begin{IEEEbiography}{Keith Underwood} {\,} is a Sr. Distinguished Technologist at HPE responsible for Slingshot NIC architecture. He co-authored the Ultra Ethernet Transport specification. Previously at Intel, he was part of the founding architecture team for OmniPath.
\end{IEEEbiography}

\begin{IEEEbiography}{Adrian Caulfield} {\,} is a Partner Engineering Manager at Microsoft, responsible for AI System and Network architecture.  Caulfield received his PhD at University of California, San Diego. Contact him at acaulfie@microsoft.com
\end{IEEEbiography}

\begin{IEEEbiography}{Abdul Kabbani} {\,} is an engineer at Microsoft and an adjunct professor at UCSC. He received his PhD at Stanford University. Contact him at abdulkabbani@microsoft.com
\end{IEEEbiography}

\begin{IEEEbiography}{Amirreza Rastegari} {\,} is the Lead HPC Performance Engineer at Microsoft Azure, where he conducts performance assessment and analysis of next-generation Azure supercomputers. He holds a PhD in scientific computing from the University of Michigan, Ann Arbor. Contact him at arastegari@microsoft.com
\end{IEEEbiography}

\end{document}